\documentclass[useAMS,usegraphicx]{mn2e}
\usepackage{rotate}
\usepackage{times}
\newif\ifAMStwofonts
\AMStwofontstrue

\def\gs{\mathrel{\hbox{\rlap{\hbox{\lower4pt\hbox{$\sim$}}}\hbox{$>$}}}}
\def\ls{\mathrel{\hbox{\rlap{\hbox{\lower4pt\hbox{$\sim$}}}\hbox{$<$}}}}

\def\asca{{\it ASCA}}

\def\xmm{{\it XMM-Newton}}

\def\et{{et al.\ }}

\def\iz{{I~Zw~1}}
\def\3c{{3C~273}}

\def\rg{{\thinspace r_{\rm g}}}
\def\fvar{{F_{\rm var}}}
\def\chidof{{\chi^2_\nu/{\rm dof}}}
\def\delchi{{\Delta\chi^2}}
\def\ka{{K$\alpha$}}

\def\feii{{Fe~\textsc{ii}}}
\def\nh{{N_{\rm H}}}

\def\deg{^{\circ}}

\def\cm{{\rm\thinspace cm}}
\def\erg{{\rm\thinspace erg}}
\def\eV{{\rm\thinspace eV}}

\def\keV{{\rm\thinspace keV}}

\def\s{{\rm\thinspace s}}
\def\ks{{\rm\thinspace ks}}

\def\cmps{\hbox{$\cm\s^{-1}\,$}}

\def\ergpscmps{\hbox{$\erg\cm^{-2}\s^{-1}\,$}}

\def\ergps{\hbox{$\erg\s^{-1}\,$}}

\def\pscm{\hbox{$\cm^{-2}\,$}}

\title[A longer \xmm\ look at \iz]
      {
A longer \xmm\ look at I Zwicky 1: Variability of the X-ray continuum, absorption,
and iron \ka\ line
      }

\author[L. C. Gallo et al.]
       {L. C. Gallo,$^{1,2}$  
        W. N. Brandt,$^3$ 
        E. Costantini,$^{4,5}$ 
	A. C. Fabian,$^6$  
        K. Iwasawa,$^1$ 
\newauthor
        and I. E. Papadakis$^7$ \\
$^{1}$ Max-Planck-Institut f\"ur extraterrestrische Physik, Postfach 1312, 85741 Garching, Germany \\
$^{2}$ SUPA, School of Physics and Astronomy, University of St Andrews, North Haugh, St Andrews, Fife KY16 9SS \\
$^{3}$ Department of Astronomy and Astrophysics, The Pennsylvania State University, 525 Davey Lab, University Park, PA 16802, USA \\
$^{4}$ SRON National Institute for Space Research Sorbonnelaan 2, 3584 CA Utrecht, The Netherlands \\
$^{5}$ Astronomical Institute, Utrecht University, P.O. Box 80000, 3508 TA
Utrecht, The Netherlands \\
$^{6}$ Institute of Astronomy, University of Cambridge, Madingley Road, Cambridge CB3 0HA\\
$^{7}$ Physics Department, University of Crete, P.O. Box 2208, 710 03 Heraklion, Crete, Greece \\
}
\date{Accepted. Received. }
\pagerange{\pageref{firstpage}--\pageref{lastpage}}
\pubyear{2007}
\begin{document}
\maketitle
\label{firstpage}

\begin{abstract}
We present the second \xmm\ observation ($85\ks$) of the
narrow-line Seyfert 1 galaxy (NLS1) \iz\
and describe its mean spectral and timing characteristics. 
On average, \iz\ is $\sim35$ per cent dimmer in 2005 than in the shorter
($20\ks$) 2002 observation.
Between the two epochs the intrinsic absorption column density diminished,
but there were also subtle changes in the continuum shape.  Considering
the blurred ionised reflection model, the long-term changes can be associated
with a varying contribution of the power law component relative to the total spectrum.
Examination of 
normalised light curves indicates that the high-energy variations are quite
structured and that there are delays, but only in some parts of the light curve.  
Interestingly, a hard X-ray lag first appears during the most-distinct structure in
the mean light curve, a flux dip $\sim25\ks$ into the observation.
The previously discovered broad, ionised Fe~\ka\ line shows significant variations 
over the course of the 2005 observation.  The amplitude of the variations is 
$25-45$ per cent and they are unlikely due to changes in the Fe~\ka-producing region, 
but perhaps
arise from orbital motion around the black hole or obscuration in the
broad iron line-emitting region.
The 2002 data are re-examined for variability of the Fe~\ka\ line at that 
epoch.  There is evidence of energy and flux variations that are associated 
with a hard X-ray flare that occurred during that observation. 
\end{abstract}



\section{Introduction}
\label{sect:intro}

I Zwicky 1 (\iz; $z=0.0611$) is considered the prototypical narrow-line Seyfert 1 galaxy (NLS1) mainly 
due to its distinct optical properties.  At X-ray energies the active galactic
nucleus (AGN) exhibits behaviour that is somewhat more modest than for typical NLS1s 
(e.g. Leighly 1999a, 1999b).  However, a short ($20\ks$) \xmm\ observation conducted 
in 2002 revealed a plethora of interesting characteristics, such as: hard X-ray
flaring, unresolved transition array (UTA) absorption, and possibly multiple 
Fe~\ka\ emission lines (Gallo \et 2004a, hereafter G04).

The discoveries in G04 prompted further examination with a second longer \xmm\ 
observation.  This is the first of three papers which describe the X-ray
properties of \iz\ utilising all \xmm\ data.  
In this work, we present an $85\ks$ \xmm\
observation and describe the mean spectral and timing characteristics of the AGN,
as well as compare the data to the earlier 2002 observation.
The complex ionised absorber and its long-term variability is discussed in detail in 
the work of Costantini \et (2007). 
In addition, \iz\ exhibits remarkable short-term spectral variability
during the 2005 observation.  This is presented in Gallo \et (2007).

\section{Observations and data reduction}
\label{sect:data}
\iz\ was observed for the second time by \xmm\ (Jansen \et 2001) on 2005
July 18--19 during revolution 1027 (obsid 0300470101)
(hereafter we will refer to this as the 2005 observation).
The total duration was $85\ks$, during which time all instruments were
functioning normally.  
The EPIC pn (Str\"uder \et 2001) and MOS (MOS1 and MOS2;
Turner \et 2001) cameras were operated in small-window mode
with the medium filter in place.
The Reflection Grating Spectrometers (RGS1 and RGS2; den Herder \et 2001)
also collected data during this time, as did the Optical Monitor
(OM; Mason \et 2001).

The Observation Data Files (ODFs) from the 2002 and 2005 observations
were processed to produce calibrated event lists using the \xmm\ 
Science Analysis System ({\tt SAS v7.0.0}).
Unwanted hot, dead, or flickering pixels were removed as were events due to
electronic noise.  Event energies were corrected for charge-transfer
inefficiencies.  EPIC response matrices were generated using the {\tt SAS}
tasks {\tt ARFGEN} and {\tt RMFGEN}.  Light curves were extracted from these
event lists to search for periods of high background flaring.
Some background flaring is detected toward the end of the 2005 observation,
and the data during those periods have been neglected.
The total amount of good
exposure time selected was $58\ks$ and $82\ks$ for the pn and MOS detectors,
respectively.
Source photons were extracted from a circular region 35$^{\prime\prime}$ across
and centred on the source.
The background photons were extracted from a larger off-source region and
appropriately scaled to the source-selection region.
Single and double events were selected for the pn detector, and
single-quadruple events were selected for the MOS.  The data quality flag was
set to zero (i.e. events next to a CCD edge or bad pixel were omitted).
The total number of source counts collected by the pn instrument in the
$0.3-10\keV$ range was approximately $2.75\times10^{5}$.  In the $0.5-10\keV$
band, the MOS instruments collected about $8.4\times10^{4}$ (MOS1) and
$8.9\times10^{4}$ (MOS2) source counts.
Comparing the source and
background spectra we found that the spectra are source dominated 
between $0.2-12\keV$.
The EPIC pn spectrum is analysed in the $0.3-10\keV$ band where the calibration is most
certain.  
To avoid cross-calibration uncertainties
at low energies, the MOS spectra are neglected below $0.5\keV$.
The 2005 spectra from all instruments show relatively good agreement in these energy
bands, within known calibration uncertainties (Kirsch 2006), so they are
fitted simultaneously unless stated otherwise.
As the broad-band light curve is less sensitive to calibration uncertainties,
the data between $0.2-12\keV$ are utilised.  

The RGS were operated 
in standard Spectro+Q mode in 2005, and a total exposure of $85\ks$ was collected. 
The RGS data are used in the analysis of the ionised absorber in \iz\ (Costantini \et
2007) and will not be discussed here.

During 2005, \iz\ was observed with the OM in Fast-mode through four filters
($U$, $UVW1$, $UVW2$ and $UVM2$).  The purpose was to examine possible optical
variability.  Unfortunately, the AGN was not well centred in the small, fast-mode 
frame and fell close to the detector edge.  Consequently, 
optical variability could not be examined in a robust manner.


\section{2005 X-ray light curves}
\label{sect:xlc}

During the $20\ks$ 2002 observation \iz\ exhibited a modest X-ray flare, which
lasted a total of $\sim5\ks$ (G04).  Notably the flare appeared to
be quite hard, concentrated in the $3-12\keV$ band, thus inducing spectral
variability.  Aside from this lone event the light curve was relatively quiet
in comparison to many NLS1.

\begin{figure}
\rotatebox{270}
{\scalebox{0.32}{\includegraphics{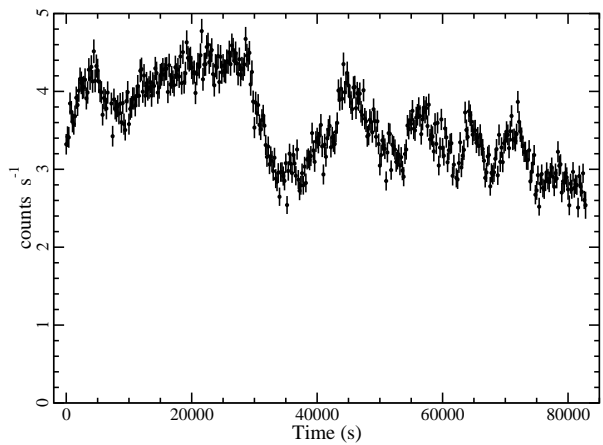}}}
\caption{The 2005 $0.2-12\keV$ EPIC pn light curve in $200\s$ bins.  
The start time corresponds to the beginning of the pn exposure.
}
\label{fig:xuvlc}
\end{figure}

The 2005 light curve appears more active, consistently variable with, on average,
larger amplitudes than observed in 2002 (Figure~\ref{fig:xuvlc}).  Perhaps the
most interesting event in the light curve is the sharp drop about $25\ks$
into the observation.
There is spectral variability related to the light curve, and
this is discussed at length in Gallo \et (2007).  Here, we will focus on the
flux variations.

\begin{figure}
\rotatebox{0}
{\scalebox{0.32}{\includegraphics{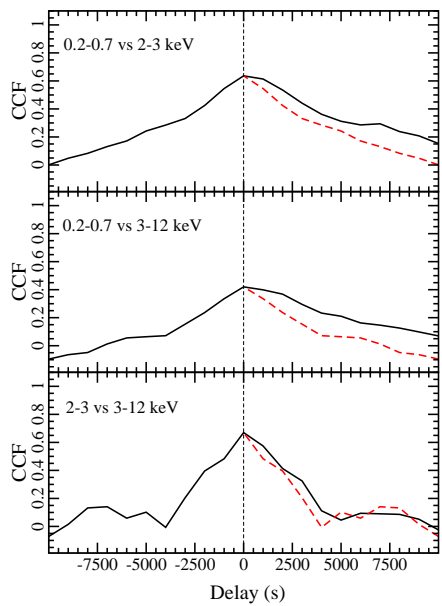}}}
\caption{The CCF from cross-correlating the energy bands shown in each panel.
In each panel the negative delay curve is reflected about the zero axis and
shown as a red, dashed curve on the positive side.  This is done to highlight
the asymmetry of the CCFs toward positive lags meaning that the soft bands
are leading the variations.
}
\label{fig:ccf}
\end{figure}

Light curves were created in three EPIC pn energy bands: $0.2-0.7\keV$, $2-3\keV$ and
$3-12\keV$.  
Using the light curves in $1000\s$ bins, each of the high-energy light curves was
cross-correlated (following Edelson \& Krolik 1988) with the the 
$0.2-0.7\keV$ band to search for possible lags
(positive delay means hard-band lags the soft-band).
All of the cross-correlation functions (CCFs) were asymmetric toward positive lags
(Figure~\ref{fig:ccf}).
As will be demonstrated in the spectral analysis, the softest band
is the region where both the power law component and the soft excess component
influence the spectrum.  Both higher bands are dominated by the power law 
component alone.

There are a few AGN where time delays as a function of Fourier frequency 
(see Nowak \et 1999 and references therein for details) have
been determined (see Papadakis \et 2001 for the first reported detection in an AGN).
If there were a {\it single} delay between the light
curves in the various energy bands, then the time delay between all of the Fourier
frequencies would be the same.  In contrast, what is typically observed
is that as the frequency of the Fourier components decreases, the delays
increase.  Furthermore, as the energy band separation between two light
curves increases, the delays, at the same frequency, also increase (e.g.
see Ar\'evalo \et 2006 for a recent discussion).

In the case of \iz, we cannot estimate time lags as a function of frequency
due to limited signal in the hard-band light curve.
However, the CCFs are consistent with what we observe in other AGN and NLS1.
The CCFs show broad humps around zero time lag with an
asymmetry toward positive lags, but not a single, distinct delay.
This is consistent with the hypothesis that in \iz, just like in other
Seyferts, the sine and cosine functions in the hard-band light curves are
delayed with respect to those in the soft-band light curves, but the delay
is not the same for all of them: the delays increase with decreasing frequency
and increased separation between the energy bands.

\begin{figure}
\rotatebox{270}
{\scalebox{0.32}{\includegraphics{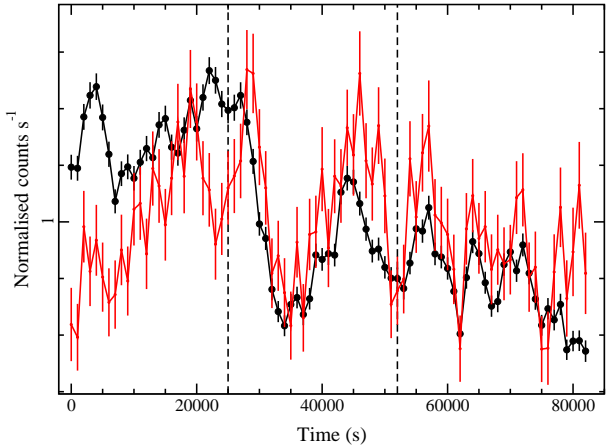}}}
\caption{Light curves in the $0.2-0.7\keV$ (black dots) and $3-12\keV$ bands
normalised to their respective means.  A hard band lag is apparent in between
$\sim25-52\ks$ (marked by the vertical dashed lines).  
The start time corresponds to the beginning of the pn exposure.
}
\label{fig:nrmlc}
\end{figure}

\begin{figure}
\rotatebox{270}
{\scalebox{0.32}{\includegraphics{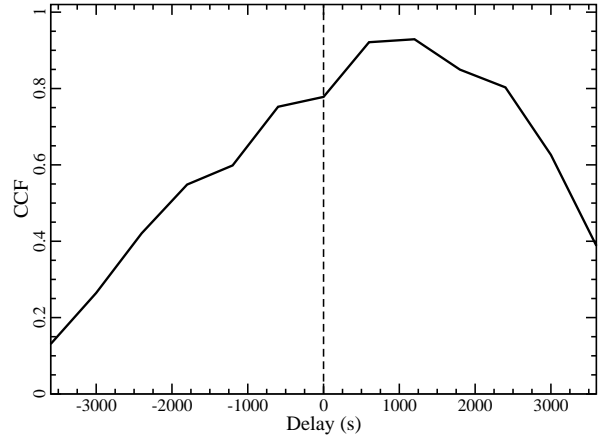}}}
\caption{The CCF when cross-correlating the $0.2-0.7\keV$ and $3-12\keV$ light
curves between $25-55\ks$ only.  A hard band lag appears more dominant in this
time band and peaks at $\sim1\ks$.
}
\label{fig:lag}
\end{figure}

Of interest may be the particularly small values measured in the CCFs 
($0.5-0.7$).
We examine this further by directly comparing light curves in each
energy band normalised to their respective mean (Figure~\ref{fig:nrmlc}).
Most obvious in the comparison between the $0.2-0.7\keV$ and $3-12\keV$ band is
the significance of the delay between $\sim25-52\ks$ (vertical dashed lines in Fig.~\ref{fig:nrmlc}).  
In particular, the delay
seemingly begins at the onset of the sharp dip in the light curve.
If we calculate the CCF in this time regime alone we see a clear dominant lag
at $\sim1\ks$ (Figure~\ref{fig:lag}).  

However, as Figure~\ref{fig:nrmlc} 
demonstrates, the lag is not always present.  Sometimes there are no
delays, and at other times the light curves show different behaviour and
even anti-correlations (e.g. $\sim20-25\ks$ and $>75\ks$ into the observation).
All of these factors contribute to the small value of the CCF peaks.
Similar behaviour was seen in the NLS1 IRAS~13224--3809 (Gallo \et 2004b)
where the authors suggested possible alternating lags between high- and low-energy
bands (i.e. sometimes hard band lags and other times it leads).

\section{2005 spectral modelling}

The source spectra were grouped such that each bin contained at least 20
counts (typically more). Spectral fitting was performed using {\tt XSPEC v11.3.0} 
(Arnaud 1996).
All parameters are reported in the rest frame of the source unless specified
otherwise.  
The quoted errors on the model parameters correspond to a 90\% confidence
level for one interesting parameter (i.e. a $\Delta\chi^2$ = 2.7 criterion).
A value for the Galactic column density toward \iz\ of
$5.07 \times 10^{20}\pscm$ (Elvis \et 1989) is adopted in all of the
spectral fits.  K-corrected luminosities are calculated using a
Hubble constant of $H_0$=$\rm 70\ km\ s^{-1}\ Mpc^{-1}$ and
a standard flat cosmology with $\Omega_{M}$ = 0.3 and $\Omega_\Lambda$ = 0.7.

In the analysis of the 2002 data, G04 were able to fit the spectra with various 
phenomenological models, but could not distinguish a superior one.  While they
were able to establish that a second, soft component was a necessary part
of the continuum they were unable to determine whether the soft excess was
curved (e.g. blackbody-like) or flat (power law-like).  

\subsection{The 2-10\keV spectra}
As was done for the 2002 data (G04), we built-up a broad-band spectral 
model for the 2005 observation utilising all the EPIC data. Fitting the
$2-10\keV$ EPIC spectra with a single power law (each instrument having
free photon index ($\Gamma$) and normalisation to accommodate existing
cross-calibration uncertainties) resulted in a satisfactory model
($\chidof=1.03/1090$).  Adding a Gaussian profile to this continuum, with
a common energy ($E$) and width ($\sigma$) among all three instruments was a 
significant improvement ($\delchi=59.5$ for 5 free additional parameters).
The best fit line parameters were $E=6.70\pm0.14\keV$, $\sigma=523^{+276}_{-157}\eV$
and $EW_{pn}=265^{+105}_{-32}$, $EW_{M1}=220^{+130}_{-49}$, 
$EW_{M2}=239^{+191}_{-79}\eV$.  
G04 introduced the possibility of multiple line features in the Fe~\ka\ region of the
2002 spectrum, including an intrinsically narrow $6.4\keV$ line.  No such line is present
in the 2005 spectrum.  A narrow, neutral line is completely insignificant in the
pn spectrum and any such feature in the MOS has an $EW<27\eV$.

A comparable improvement ($\delchi=38.0$
for 5 free additional parameters) can be achieved if a
disc line profile (Fabian \et 1989) rather than a Gaussian profile is used for the
broad residuals.  In this case the best-fit energy and inner disc radius are
$E=6.55^{+0.20}_{-0.10}\keV$ and $R_{in}=6-26$ gravitational radii, respectively.   
The disc inclination ($i$) is fixed at $50\deg$, to remain consistent with
our finding in Section~\ref{sect:ref05}.  Leaving the parameter
free to vary does not enhance the fit statistics, nor does it affect the best-fit
line energy. 
The line strength is $EW\approx200\eV$ in each spectrum.   
For simplicity, we adopt the Gaussian profile.

\subsection{Simple broad-band models}
\label{sect:sbbm}
Extrapolating the $2-10\keV$ fit to $0.3\keV$ in the pn and $0.5\keV$ in the two MOS
left a gradual rise in the residuals that is typically associated with the
soft excess.  
\begin{table*}
\begin{center}
\caption{Fit parameters for the simple models (column 1) applied to the 2005
full-band data.  Fit quality is given in column 2.  Neutral absorption intrinsic 
to \iz\
is given in column 3 ($\times10^{20}\pscm$).  In column 4, the continuum parameters
are shown.  If the parameter is not linked between the various instruments it is
identified as pn, M1 (MOS1) and M2 (MOS2).  $E_b$ is the break energy (in \keV) 
for the broken power law model.  The photon index above ($\Gamma_h$) and below
($\Gamma_s$) the break energy are also shown.  The blackbody temperature ($kT$)
is given in \eV.  Column 5 provides the absorption-edge parameters: energy ($E$)
and optical depth ($\tau$).  All fits except for the partial covering model
(second row) include an Fe\ka\ emission line with
comparable parameters to the results presented in the text.  The line was not
statistically necessary in the partial covering model.  The parameters of
the partial covering absorber, column density ($\nh_{pc}$, $\times10^{22}\pscm$) 
and covering fraction ($f$), are given in column 6.
}
\begin{tabular}{cccccc}                
\hline
(1) & (2) & (3) & (4) & (5) & (6) \\
Model  &  $\chidof$  &  $\nh_{i}$ & Continuum  &  Edge  & Partial absorber\\
\hline
Single   & $1.13/1628$ & $0.71^{+0.23}_{-0.20}$ & $\Gamma_{pn}=2.37\pm0.01$ & $E=644^{+9}_{-10}$  & \\
power law&           &                        & $\Gamma_{M1}=2.29^{+0.02}_{-0.01}$ & $\tau=0.15\pm0.02$ &\\
         &           &                        & $\Gamma_{M2}=2.30^{+0.02}_{-0.01}$ & &\\
\hline
Single   & $1.08/1629$ & $1.62^{+0.28}_{-0.31}$ & $\Gamma_{pn}=2.43\pm0.06$ & $E=648^{+6}_{-9}$ & $\nh_{pc}=31.5^{+7.6}_{-4.1}$\\
power law&           &                        & $\Gamma_{M1}=2.35\pm0.01$ & $\tau=0.19\pm0.02$ & $f=0.72^{+0.09}_{-0.06}$ \\
with partial covering&           &            & $\Gamma_{M2}=2.37\pm0.01$ & \\
\hline
Broken   & $1.08/1626$ & $2.13^{+0.31}_{-0.38}$ & $\Gamma_{s}=2.48\pm0.03$   & $E=648^{+6}_{-9}$ & \\
power law&           &                        & $E_b=1.89^{+0.23}_{-0.10}$ & $\tau=0.20\pm0.02$ & \\
         &           &                        & $\Gamma_{h,pn}=2.27^{+0.02}_{-0.03}$ & & \\
         &           &                        & $\Gamma_{h,M1}=2.17^{+0.03}_{-0.05}$ & & \\
         &           &                        & $\Gamma_{h,M2}=2.35\pm0.03$ & \\
\hline
Blackbody& $1.08/1626$ & $<0.20$                & $kT=216\pm12$                        & $E=651^{+7}_{-9}$ & \\
plus     &           &                        & $\Gamma_{pn}=2.30\pm0.01$          & $\tau=0.20\pm0.02$ & \\
power law&           &                        & $\Gamma_{M1}=2.21\pm0.01$          & & \\
         &           &                        & $\Gamma_{M2}=2.23\pm0.01$          & & \\
\hline
Double   & $1.04/1626$ & $3.22^{+0.69}_{-0.27}$&$\Gamma_{1}=2.67^{+0.05}_{-0.04}$    & $E=652^{+6}_{-10}$ & \\
power law&           &                        & $\Gamma_{2,pn}=1.43^{+0.16}_{-0.06}$ & $\tau=0.18\pm0.02$ & \\
         &           &                        & $\Gamma_{2,M1}=1.74^{+0.12}_{-0.04}$ & & \\
         &           &                        & $\Gamma_{2,M2}=2.03^{+0.13}_{-0.06}$ & & \\
\hline
\label{tab:fits}
\end{tabular}
\end{center}
\end{table*}

\begin{figure}
\rotatebox{270}
{\scalebox{0.32}{\includegraphics{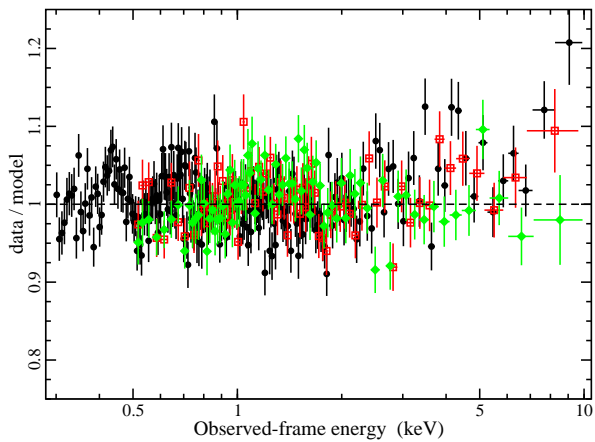}}}
\caption{The residuals (data/model) from fitting a model composed
of a blackbody, power law and Gaussian profile 
(modified by absorption) to the EPIC data (see third row in
Table~\ref{tab:fits}).  
Black dots, red open squares, and green diamonds
correspond to the pn, MOS1, and MOS2 data, respectively.
The data are rebinned for display purposes. 
}
\label{fig:epicres}
\end{figure}
The entire 2005 $0.3-10\keV$ band was modelled with various continuum
models.  Through repeated effort we found it necessary to include in all
models an additional
neutral absorber at the redshift of the source, as well as
a low-energy ($\sim650\eV$) absorption edge that mimics the 
more complicated ionised absorption.

As with the 2002 observation, the 2005 data show that a second continuum component 
is required to fit the spectrum of \iz\ (see Table~\ref{tab:fits}).
According to an $F$-test, the addition of a second 
continuum component (i.e. 2 additional parameters) to a single power law is highly 
significant ($F$-test probability $\sim10^{-17}$).  
Also like the 2002 observation, a single superior model does not stand out.
However, there are similarities among all the models in Table~\ref{tab:fits}. 
One may notice that the second component is not highly curved.  
Even in the case of the blackbody plus power law model, the high-temperature
of the blackbody component will give it a flatter appearance in the EPIC
bandpass.

Of interest is that there is detectable diminishing of the intrinsic 
column density in \iz\ since 2002.  The exact change 
depends on what continuum model is assumed, but appears to be at least on
the $50$ per cent level.

The residuals from the blackbody plus power law fit
are shown in Figure~\ref{fig:epicres}.
Notable in Figure~\ref{fig:epicres} are the positive residuals, which remain
in the fit above $\sim8\keV$.  These residuals are apparent in all 
broad-band fits to the 2005 data (see also Figure~\ref{fig:reffit} and
the lower panel of Figure~\ref{fig:pnreffit}) , with the exception of the double power
law model.  The most likely cause is the significant spectral variability exhibited
by \iz\ in 2005 (Gallo \et 2007).  Gallo \et suggest that the power law
photon index could be changing during the 2005 observation, which would explain the 
difficulty in fitting the average spectrum with a single power law.
The reason that the residuals are minimal in the double power law model
could simply be that the second power law does not begin to dominate the spectrum
until rather high energies ($\sim5.5\keV$ in the pn spectrum).

Based on the blackbody plus power law model, the average $0.3-10\keV$ flux
during 2005, corrected for Galactic absorption, is $\sim1.4\times10^{-11}\ergpscmps$.
The luminosities in the $0.3-10\keV$ and $2-10\keV$ bands are 
$1.28\times10^{44}\ergps$ and $0.43\times10^{44}\ergps$, respectively.  \iz\ is
$\sim35$ per cent dimmer than it was in 2002. 
For comparison, the estimated $2-10\keV$ luminosity during the \asca\ observation
was $0.53\times10^{44}\ergps$ (Leighly 1999b).

\section{A comparison of the 2002 and 2005 spectra}

\subsection{Summary of the 2002 observation}
The $0.2-10\keV$ spectrum of \iz\ has long been known to differ from those of
other NLS1.  It possesses a rather weak soft excess and is modified by intrinsic
neutral and ionised absorption.  The 2002 observation of \iz\ (G04) was rather
typical of what had been seen previously (e.g. Leighly 1999a, b).

There were several simple continuum models (e.g. blackbody plus power law,
broken power laws, multiple blackbodies plus power law) that when modified
by intrinsic ionised absorption
(an absorption-edge),
neutral absorption (absorption column density) and an Fe~\ka\ emission line,
fit the EPIC spectra adequately in 2002.  None of these continuum models
stood out as superior.
Since the 2002 data were reprocessed with the newest calibration
we refit the simple models to the newly processed data and find comparable
results.  The one modification was in the intrinsic absorption column density, 
which
was slightly lower with the new calibration ($\nh=4.5\pm0.5\times10^{20}\pscm$ 
as opposed to $\nh=10.0\pm0.1\times10^{20}\pscm$).

A particularly interesting spectral result from the 2002 observation was the
possible presence of multiple iron lines.  In addition to a broad and ionised
line, the spectrum also required a narrow, unresolved emission line at $6.4\keV$
.
The most obvious interpretation adopted by G04 was that this feature was 
produced
in distant material such as the torus, but an alternative interpretation is
offered in Sect~\ref{sect:dis02}.

\subsection{Spectral changes since 2002}
\label{sect:spchg}
\begin{figure}
\rotatebox{270}
{\scalebox{0.32}{\includegraphics{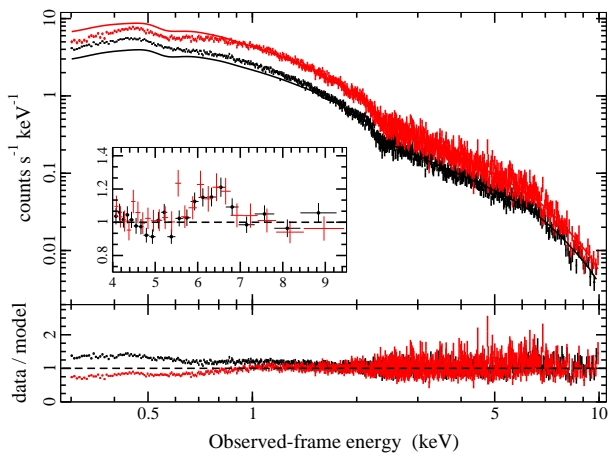}}}
\caption{Top panel:
Best-fit power law modelled to the 2002 (red marks, $\Gamma=2.28\pm0.03$)
and 2005 (black marks, $\Gamma=2.15\pm0.02$) $2-10\keV$ pn spectrum, and
extrapolated to $0.3\keV$.  Lower panel: Resulting ratio from the above model.
Top panel (inset): Ratio in the $4-9.5\keV$ band, binned for display
purposes.
}
\label{fig:po}
\end{figure}

The 2005 spectrum does appear substantially different from the 2002 observation.
The broad-band differences between 2002 and 2005
can be depicted by extrapolating the best-fit $2-10\keV$ power law to both 
spectra
down to lower energies (Figure~\ref{fig:po}).

The 2005 observation shows a much softer spectrum, a suggestion that the
level of absorption may have changed since 2002.  The best-fitting power law in the
$2-10\keV$ band is slightly flatter in 2005 ($\Gamma=2.15\pm0.02$ compared to
$\Gamma=2.28\pm0.03$ in 2002), which likely also contributes to the apparently
stronger soft excess at that epoch.  The inset in Figure~\ref{fig:po} shows
the residuals in the Fe~\ka\ band.  As with previous observations, the iron line
appears broad and ionised.  The agreement in the appearance of the iron line
at both epochs is good.

We applied the blackbody plus power law model from 2002 to the 2005 data to 
examine consistency between
the two epochs.  Simply rescaling the model was unacceptable ($\chidof=5.7/846$).
Allowing the level of neutral absorption to vary freely significantly improves
the fit quality ($\chidof=1.27/845$) indicating that the column density
of the neutral absorber has dropped in 2005 to $\nh\approx1.4\times10^{20}\pscm$.
Allowing the low-energy absorption edge to vary in energy and depth is a
further improvement ($\delchi=36.2$ for 2 additional free parameters).
The overall
fit is not acceptable ($\chidof=1.23/843$)
and demonstrates that subtle changes
in the intrinsic continuum shape since 2002 may be appropriate.

If rather we consider a constant absorber and allow only the continuum parameters
to vary, we obtain a worse fit ($\chidof=1.35/843$).  Indeed variable absorbers
seem necessary and are supported by the RGS data
(Costantini \et 2007).

\section{A reflection interpretation for \iz}

\subsection{The 2005 observation}
\label{sect:ref05}
The spectrum of \iz\ is power law dominated, but there is the necessity for a small
contribution from a soft component below $\sim2\keV$.  The ionised disc reflection
model (Ross \& Fabian 2005) blurred by relativistic effects has had relatively
good success in duplicating the appearance of the soft excess 
(e.g. Crummy \et 2006).

All the EPIC spectra were fitted with the blurred reflection plus power
law model.  As in the previous fits, we included an intrinsic neutral absorption
component and an
absorption edge as a pseudo-ionised absorber.
Since the model is complex, to simplify the fitting process the normalisations of
the MOS spectra were linked to the pn spectrum and only a constant scale factor 
was used to correct for possible flux offsets between instruments.
The incident spectrum for the reflection component was linked to that of the
power law component.  
\begin{figure}
\rotatebox{270}
{\scalebox{0.32}{\includegraphics{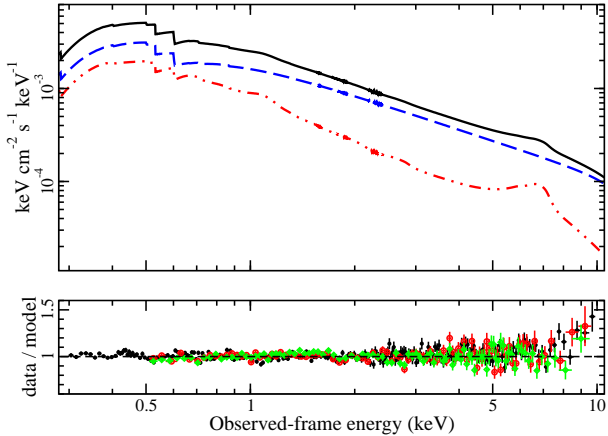}}}
\caption{Top panel: The best fit blurred reflection model (only the pn
model is shown for clarity).  The reflection (red, dash-dot curve) and
power law (blue, dashed curve) components, which make up the combined
fit (black, solid curve) are presented.  Bottom panel:  The residuals of
the model to the EPIC spectra.  Symbols are as defined in Figure~\ref{fig:epicres}.
}
\label{fig:reffit}
\end{figure}
\begin{table}
\begin{center}
\caption{The blurred reflection and power law model (with absorption) applied
to the 2005 EPIC spectra ($\chidof = 1.10/1627$). 
The absorption parameters are as defined in Table~\ref{tab:fits}.
The blurring parameters are the emissivity index ($\alpha$); inner ($R_{in}$)
and outer ($R_{out}$) disc radius in gravitational radii ($\rg=GM/c^2$); 
and disc inclination ($i$).  The iron abundance and ionisation
parameter ($\xi=4\pi Fn^{-1}$, where $F$ is the total incident flux and $n$ is
the hydrogen number density) of the reflector are also given.
}
\begin{tabular}{cc}                
\hline
(1) & (2)  \\
Model Component  &  Fit Parameter  \\
\hline
Absorption   &  $\nh_{i}=1.05^{+0.07}_{-0.13}\times10^{20}\pscm$ \\
(neutral and    &  $E=638^{+7}_{-3}\eV$ \\
 ionised)     &  $\tau=0.30\pm0.01$ \\
\hline
Power law      & $\Gamma=2.24\pm0.01$ \\
\hline
Blurring       & $\alpha=2.48^{+0.36}_{-0.31}$ \\
               & $R_{in}=5.0^{+8.2}_{-1.6}\rg$  \\
               & $R_{out}=200\rg$ (fixed) \\
               & $i=51\pm2\deg$ \\
\hline
Reflector     & Fe$=1.1\pm0.1$ solar \\
              & $\xi=3060^{+107}_{-186}\erg\cmps$ \\
\hline
\label{tab:reffit}
\end{tabular}
\end{center}
\end{table}

The fit quality is not as satisfactory as some of the simple models 
($\chidof=1.10/1627$), but it does have the advantage of offering a physical
explanation for the origin of the soft excess and broad Fe~\ka\ line.  
Fit parameters
are shown in Table~\ref{tab:reffit}, and the model along with fit residuals are
presented in Figure~\ref{fig:reffit}.  The fraction of reflected-to-total flux
in the $0.3-10\keV$ range (corrected for Galactic and source absorption) is 
$\sim0.30$.
Drawing conclusions from the fit parameters should probably be done with caution 
given the complexity of the low-energy absorption.  However, it can be noted that
the disc inner radius is not extremely small ($R_{in}=5.0^{+8.2}_{-1.6}\rg$) and
the emissivity profile is not very steep ($\alpha=2.48^{+0.36}_{-0.31}$).  
Both points suggest that the reflection component does not necessarily 
originate from the innermost confines around the black hole as is suggested for
many NLS1. 

The advantage of this model is that it can account for the
iron emission and soft excess in a self-consistent manner.  Moreover, 
as will be discussed
in Gallo \et (2007), this reflection scenario can potentially explain some of the
complicated spectral variability exhibited by \iz. 

\subsection{Simultaneous fit to the 2002 and 2005 data}

Based on the investigation in Section~\ref{sect:spchg}, most of the spectral
differences between 2002 and 2005 can be associated with changes in the level of
absorption.  However, there do appear to be subtle changes in the shape of the
continuum as well.  As \iz\ was nearly twice as bright in 2002, in terms of
the blurred reflection model the changes in the continuum shape could be
interpreted as differing contributions from the direct power law component.

\begin{figure}
\rotatebox{270}
{\scalebox{0.32}{\includegraphics{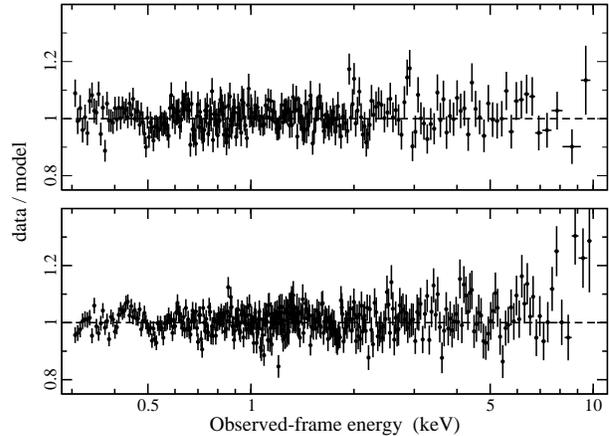}}}
\caption{Residuals from a self-consistent power law plus blurred reflection model
fitted simultaneously to the pn data from 2002 (top panel) and 2005 (bottom panel).
}
\label{fig:pnreffit}
\end{figure}

The data from both epochs were fitted simultaneously to examine consistency
in the reflection scenario.  For simplicity, many parameters were fixed to the
best-fit 2005 values (see Table~\ref{tab:pnref}).  The primary parameters that
were left free for each dataset were the power law and absorption parameters, and
the normalisation of the reflection component.  The iron abundance and disc 
inclination were linked between the two spectra as large changes in these
parameters would not be expected over 3-years.  A reasonable fit can be
achieved ($\chidof = 1.06/1480$) in this manner (see Figure~\ref{fig:pnreffit}
and Table~\ref{tab:pnref}).
\begin{table}
\begin{center}
\caption{The blurred reflection and power law model (with absorption) applied
simultaneously to the 2002 and 2005 pn spectra ($\chidof=1.06/1480$). 
The parameters are as defined in Table~\ref{tab:reffit} and $f$ marks the
ones that are fixed in the fit.  The 2002 and
2005 values are given in columns (2) and (3), respectively.  The reflection ($R$) and
flux fractions ($F_R$) (as defined in the text) are given in the final two rows.
}
\begin{tabular}{ccc}                
\hline
(1) & (2) & (3) \\
Model Component  &  2002  & 2005 \\
\hline
Absorption   &  $\nh_{i}=4.38^{+0.20}_{-0.19}$ & $0.22^{+0.14}_{-0.17}\times10^{20}\pscm$ \\
(neutral and    &  $E=623\pm10$ & $634^{+11}_{-6}\eV$ \\
 ionised)     &  $\tau=0.30^{+0.03}_{-0.04}$ & $0.23\pm0.02$ \\
\hline
Power law      & $\Gamma=2.28\pm0.02$ & $2.21^{+0.10}_{-0.08}$ \\
\hline
Blurring       & $\alpha=2.48^{f}$ & $2.48^{f}$\\
               & $R_{in}=5.0^{f}$ & $5.0^{f}\rg$  \\
               & $R_{out}=200^{f}$  & $200^{f}\rg$\\
               & $i=58\pm3$ & $58\pm3$\\
\hline
Reflector     & Fe$=1.10\pm0.13$ & $1.10\pm0.13$ solar \\
              & $\xi=3060^{f}$ & $3060^{f}\erg\cmps$ \\
\hline
$R$  & $0.27\pm0.04$ & $0.45\pm0.04$ \\
$F_R$  & $0.22\pm0.03$ & $0.31\pm0.03$ \\
\hline
\label{tab:pnref}
\end{tabular}
\end{center}
\end{table}

As is seen in Table~\ref{tab:pnref} there is no strong indication for a change
in the shape of the direct power law component between 2002 and 2005.  In addition,
most of the blurring and reflection parameters have been fixed.  The most
notable changes are in the amount of absorption (as noted in previous fits and
by Costantini \et 2007), 
the reflection fraction ($R$, fraction of reflected
emission to power law emission in the $0.3-10\keV$ band), 
and in the level at which the reflection component
contributes to the total $0.3-10\keV$ spectrum ($F_R=0.22\pm0.03$ in 2002
and $0.31\pm0.03$ in 2005).

\section{Iron line variability}
\subsection{Fe\ka\ variability during the 2005 observation}

In the analysis of G04 there were indications that the broad feature in the 2002
observation was composed of multiple (at least two) narrower emission lines.
In the 2005 observation, the residuals can be fit by a single Gaussian profile.
Based on the width of the line ($\sigma=523^{+276}_{-157}\eV$) in the EPIC spectra,
the
line-emitting region is between $12-66\rg$.  Similar distance estimates come from the 
disc line fit and the blurred reflection model.  Given the long observation
and proximity of the line emitting region makes it plausible to search for line 
variability.  

The mean pn spectrum was divided into 17 segments, each of $5\ks$.  The exposure in
each segment was $\sim3.5\ks$, taking account of the live-time of the pn camera
in small window mode. 
The average number of source counts in the
$5.0-7.2\keV$ band was slightly more than $300$.  The counts were grouped into spectra 
such
that each bin contained at least 20 counts. 
For each $5\ks$ spectrum the continuum was determined by fitting the $4-10\keV$
band with a power law, while excluding the $5-7.2\keV$ range.
The omitted band was replaced and the spectrum was rebinned into channels of
$250\eV$.  
Excess counts above the continuum in each
$250\eV$ bin were recorded.  From this information an excess residual image
was created as illustrated in Iwasawa \et (2004; and in prep).
The result is analogous to modelling each spectrum with a power law plus 
Gaussian profile and measuring the flux in the Gaussian component.  The 
advantage of creating an excess residual image is that we can circumvent
the limitations of working in the Poisson regime, thus making trends in the
variations of the line-flux more visible.
\begin{figure}
\rotatebox{0}
{\scalebox{0.32}{\includegraphics{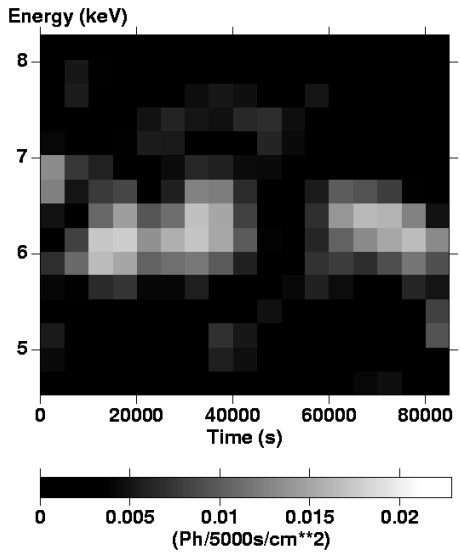}}}
\caption{Excess-residual image for the 2005 pn spectrum in the $4.5-8.2\keV$ band.  
Bright spots indicate positive residuals in the spectrum, above a power law
continuum.  The time resolution is $5\ks$, and the spectra are binned into
$250\eV$ wide channels.  The bright band between $5.8-6.7\keV$ corresponds
to the broad emission line seen in the mean spectrum.  The feature appears
to be variable over the course of the observation.
}
\label{fig:6kevex}
\end{figure}

The resulting image from the 2005 observation is shown in Figure~\ref{fig:6kevex}
and clearly indicates that the broad-line flux is variable.  Most notably, the
line flux diminishes significantly $45-60\ks$ from the start of the observation.

To create a light curve from the image, the excess counts between
$5.8-6.7\keV$ were accumulated.  Uncertainties on the light-curve data points were
estimated from Monte Carlo simulations of the excess residual image.
\begin{figure}
\rotatebox{270}
{\scalebox{0.32}{\includegraphics{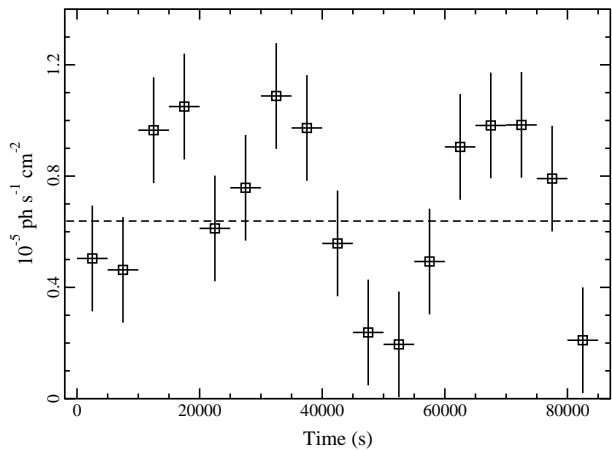}}}
\caption{The light curve of the broad Fe~\ka\ emission in \iz\ during the
2005 observation.  Bins are $5\ks$.  The 1-$\sigma$ uncertainties
were estimated as described in the text.
A constant fit is shown for comparison.
}
\label{fig:6kevlc}
\end{figure}
One-thousand images were created in an identical manner as done for \iz.
The spectra from which the images were derived were
simulated for identical conditions as the 2005
observation, assuming a power law plus stationary Gaussian profile.
The uncertainty in the observed light curve was taken to be
the standard deviation in the excess counts resulting from the simulated images.
Details of the process including caveats are discussed in Iwasawa \et (2004).

The Fe~\ka\ light curve shows clear variability (Figure~\ref{fig:6kevlc})
with a constant fit being unacceptable ($\chi^2_{\nu}=2.45$).
The fractional variability amplitude ($\fvar$, using uncertainties derived
in Edelson \et 2002) is $34.8\pm9.7$ per cent.

\begin{figure}
\rotatebox{270}
{\scalebox{0.32}{\includegraphics{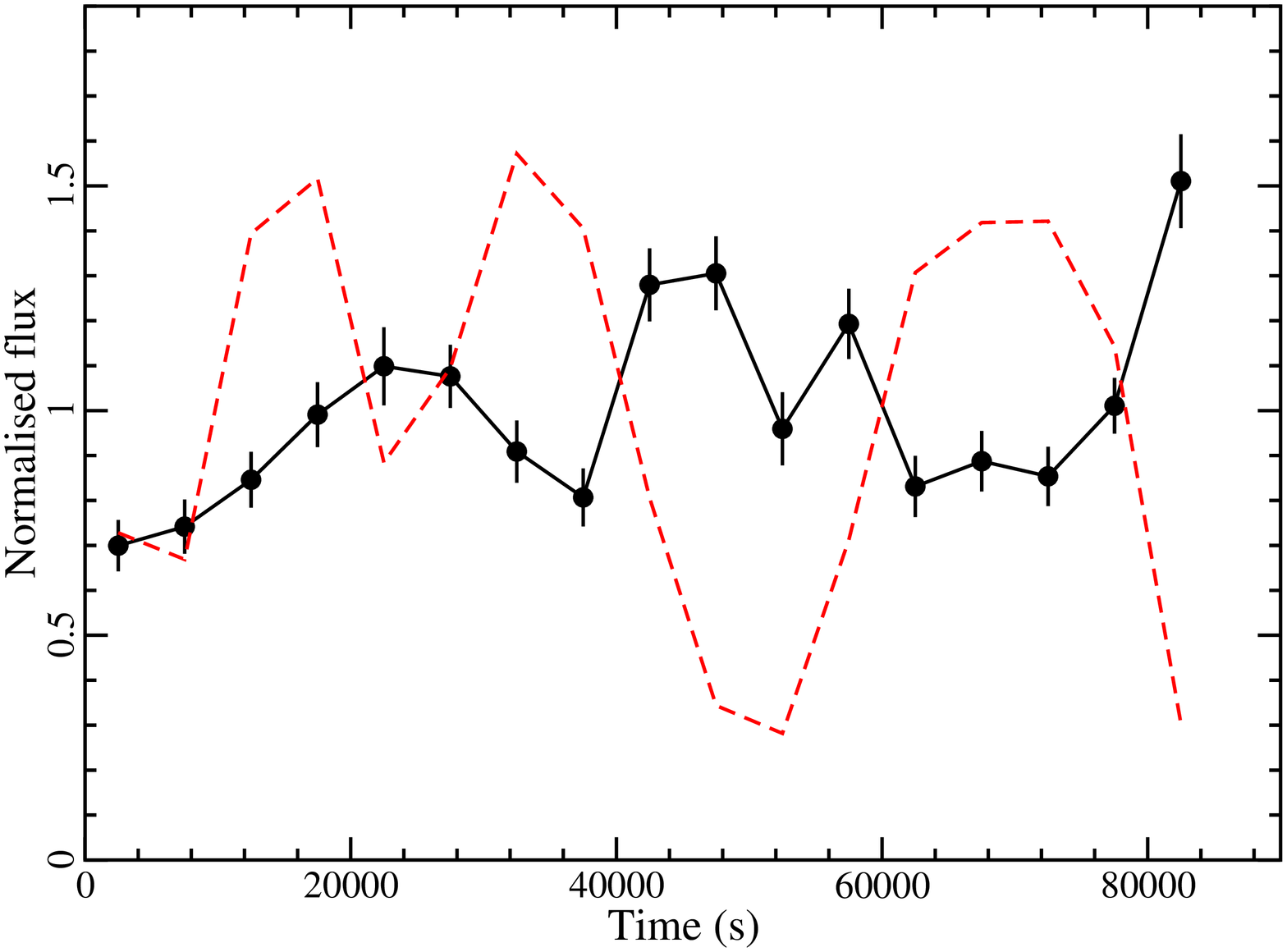}}
\scalebox{0.32}{\includegraphics{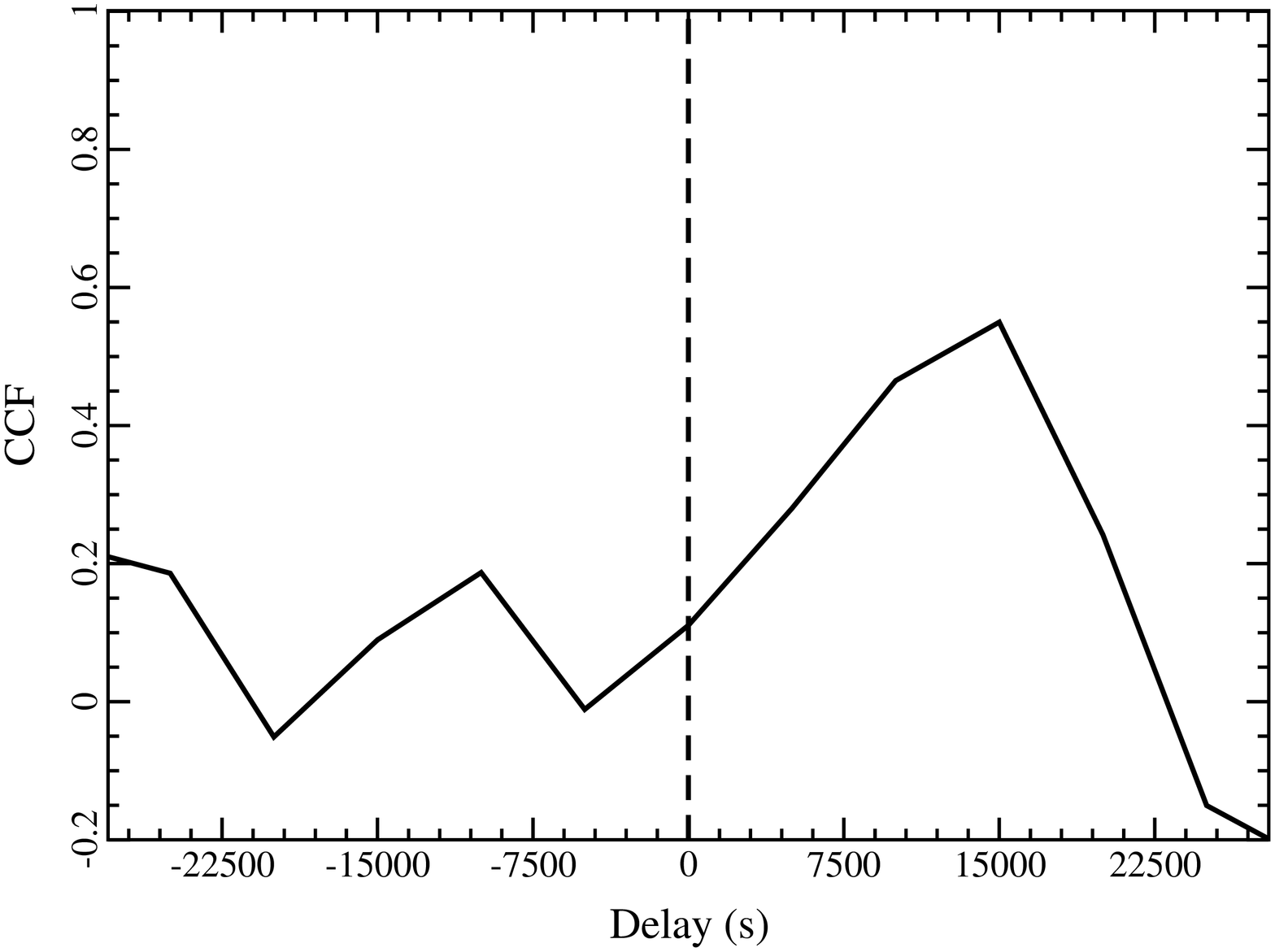}}}
\caption{Top panel: The $6-10\keV$ light curve of the continuum used
to establish the iron line variations, normalised to its average flux.
The Fe\ka\ data shown in Figure~\ref{fig:6kevlc} are normalised to the 
average Fe\ka\ flux and overplotted
(red, dashed curve).
Bins are $5\ks$.  The 1-$\sigma$ uncertainties on the continuum data
points are taken from the $6-10\keV$ count rate.
Lower panel: The CCF of the two light curves shown in the top panel.
A positive delay indicates that the continuum changes lead the emission line
variations.
}
\label{fig:cline}
\end{figure}

A light curve of the $6-10\keV$ continuum was generated based on the power law fits
used to derive the Fe~\ka\ light curve.  The uncertainties in the $6-10\keV$
flux were assumed to be the same as the uncertainty in the corresponding count rate.
The continuum light curve, normalised to its average flux, is shown 
in Figure~\ref{fig:cline} 
with the broad-line flux light curve
(Figure~\ref{fig:6kevlc}), normalised to its average flux, 
overplotted as a dashed curve.
Interestingly, the curves appear to be anti-correlated in some parts of
the light curve.  Cross-correlating the broad-line and $6-10\keV$ continuum 
light curves indicates a weak correlation with the broad-line light curve lagging 
the continuum fluctuations by $\sim15\ks$ (lower panel Figure~\ref{fig:cline}).

The fractional variability in the $6-10\keV$ continuum is $\fvar=20.8\pm4.0$
per cent, less than the $\fvar$ in the Fe~\ka\ line.
Statistically, the amplitude of the continuum variations is consistent with
the Fe~\ka\ variations within about 2-$\sigma$.  However, the difference is
likely larger since the uncertainties on $\fvar$ are conservatively
over-estimated (Edelson \et 2002).

\subsection{Revisiting the Fe~\ka\ variability in 2002}
In fitting the spectra in various segments of the 2002 observation, G04
found little evidence of intensity variations in the broad-line.
We take advantage of the excess-residual technique to re-examine   
the broad-line variability in the 2002 observation.
An excess-residual image has been created for the 2002 pn observation in a nearly
identical manner as was done for the 2005 observation.  The only difference
is that the time resolution is slightly better during 2002 because the mean 
spectrum was divided into 6 segments of $3.6\ks$ each (rather than $5\ks$).
This can be done without degrading the data quality because the \iz\ was brighter
in 2002 and because the observation
was conducted in large-window mode, which has a CCD live-time of $\sim95$ per
cent; therefore the actual exposure in each segment of the 2002 observation is
almost the same as the exposure in each segment of the 2005 data.  
\begin{figure}
\rotatebox{0}
{\scalebox{0.32}{\includegraphics{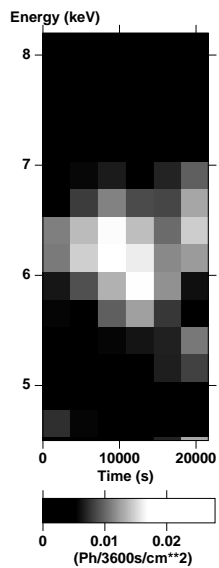}}}
\caption{Excess-residual image for the 2002 pn spectrum in the $4.5-8.2\keV$ band.  
Bright spots indicate positive residuals in the spectrum, above a power law
continuum.  The time resolution is $3.6\ks$, and the spectra are binned into
$250\eV$ wide channels.  The bright band between $6.0-6.5\keV$ corresponds
to the broad emission line seen in the mean spectrum.  Between $10-15\ks$ there
is also brightening in the $5.5-6.0\keV$ band.  
}
\label{fig:2002res}
\end{figure}
\begin{figure}
\rotatebox{270}
{\scalebox{0.32}{\includegraphics{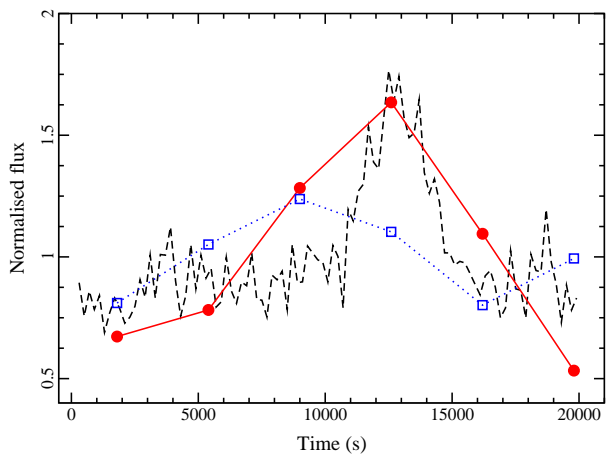}}}
\caption{Various light curves from the 2002 observation of \iz\ normalised
to their respective averages.  The black, dashed curve is the $3-12\keV$ continuum
light curve highlighting the hard X-ray flare reported previously by G04 
(uncertainties in the curve are found in G04).  The filled, red circles connected
by a solid line mark the light curve extracted from the residual image
(Figure~\ref{fig:2002res}) in the $5.5-6.0\keV$ range (i.e. the red line).
The peak intensity in the red line corresponds with the
peak intensity in the $3-12\keV$ continuum light curve.
The open, blue squares
connected by a dotted line is the light curve extracted from the residual image
in the $6.0-6.5\keV$ band (i.e. the blue line).  
}
\label{fig:cline2002}
\end{figure}

The 2002 excess-residual image is shown in Figure~\ref{fig:2002res}.
The broad-line is dominant in the $6.0-6.5\keV$ range, but there 
is also apparent brightening that occurs in the $5.5-6.0\keV$
band between $10-15\ks$.  
This is of particular interest because the continuum light curve shows
a hard X-ray flare that peaks during this time interval (see G04).

Light curves were extracted from the excess-residual 
image in the $6.0-6.5\keV$ (blue line) and $5.5-6.0\keV$ (red line)
bands.  They were then normalised
to their respective means and plotted with the normalised 2002 $3-12\keV$ continuum 
light curve (Figure~\ref{fig:cline2002}).  As the observation was so
short uncertainties in the residual-image light curves were not derived
in a robust manner as was done for the 2005 data.  
We also note that propagation of Poisson uncertainties in each data bin yields
very large error bars (on the order of 100 per cent), therefore we are only 
illustrating a trend in Figure~\ref{fig:cline2002}.

The light curves suggest that the variations are more intense in the 
$5.5-6.0\keV$-line than in the $6.0-6.5\keV$-line.
There also is no obvious connection between
the $6.0-6.5\keV$-line light curve and that of the continuum.
On the other hand, the $5.5-6.0\keV$ variations appear to be well correlated with the
high-energy continuum variations and, despite the unknown size of the true error bars
in the $5.5-6.0\keV$-line light curve, the trend is obvious.

\section{Discussion}

\subsection{\iz: a not so prototypical NLS1 in X-rays?}
The strong \feii\ emission in the optical spectrum of \iz\ was noted a 
half century ago (Sargent 1968), and given the intensity and narrowness of
the lines
the AGN became a laboratory for \feii\ studies (e.g. Phillips 1976, 1977, 1978;
Boroson \& Green 1992; Laor \et 1997; Vestergaard \& Wilkes 2001).  
Its narrow Balmer lines were also noted fairly early on 
(e.g. Oke \& Lauer 1979).

When a formal definition of the NLS1 class
was established (Osterbrock \& Pogge 1985), based on optical spectral 
properties, it was realised that \iz\ was
already the best-studied member of this class of AGN, likely leading to its 
labelling
as the prototype NLS1.
However, some X-ray properties are also
characteristic of NLS1s 
and \iz\ does not exhibit all these qualities to noticeable
levels.  In fact, in some instances \iz\ even displays behaviour that is
atypical of the class.
For example,
\iz\ is observed through a column density intrinsic to the AGN.
This is unusual for NLS1s (e.g. Boller, Brandt \& Fink 1996), but also for type
1 AGN in general.  

The soft excess emission above the underlying steep
power law is weak in \iz\ compared to most NLS1s, and not because of the 
relatively high intrinsic
absorption (e.g. Leighly 1999b, Tanaka \et 2005).  Even the clear detection
of a broad Fe~\ka\ emission line in \iz\ seems unusual given that NLS1s appear 
not to show broad emission lines so obviously (e.g. Fabian \et 2002, 2004;
Guainazzi \et 2006).

The identification of \iz\ as the prototype NLS1 originated due to its 
well-studied optical
properties, but it now lingers mainly as a historic perspective.  It can even
be argued that some of its X-ray properties are atypical of the norm.

\subsection{Time lags}
The average CCFs calculated in Section~3 show a tendency for positive lags (hard band following soft band).
This phenomenon has been observed in other AGN (e.g. Ar\'evalo \et 2006
and references therein) and Galactic black holes (e.g. McClintock \& Remillard 2005).
However the origin of the lags is not certain.

In a simple corona with a uniform temperature and optical depth, variations can
result from fluctuations in the intensity of the seed photons.  Variations of
this nature should produce fixed delays between low- and high-energy bands, 
resulting in a common time delay for all Fourier components.  We do not estimate
time lags as a function of frequency in \iz; however the broad, asymmetric CCFs
suggest that we do not have a single, well defined delay between the
various energy bands.  
Examination of normalised light curves in different energy bands further supports 
this.  A time lag is only seen in parts of the light curves. In fact, at times
the light curves are not correlated, or are actually anti-correlated.

Lags are seen in Cyg~X-1, which can be attributed to spectral evolution in
magnetic flares as they detach from the accretion disc 
(e.g. Poutanen \& Fabian 1999).  Another possibility is that the lags are
manifested by material that propagates toward the black hole from regions of soft
X-ray emission to regions of hard X-ray emission (e.g. B\"ottcher \& Liang 1999;
Kotov \et 2001).

Based on the observation of another NLS1, IRAS~13224+3809, Gallo \et (2004b) 
claimed a lag that alternated (sometimes hard band followed other times it
led), with possible dependency on the average flux of the source.
There is likely similar behaviour here as the first notable lag coincides with a
sharp flux dip in the light curve about $25\ks$ into the observation.

\subsection{Diminished neutral absorption since 2002}
Significant variations in the low-energy spectra of some Seyfert~1
galaxies have been previously observed and interpreted as changes in the
low-energy continuum component (e.g. Page \et 2002, Starling \et 2004).
In the case of \iz, this does not appear to be the most straightforward
explanation.
There is detectable diminishing of the intrinsic 
column density in \iz\ since 2002.  The exact change 
depends on what continuum model is assumed, but appears to be at least on
the $50$ per cent level.

Changes in the neutral column density have been reported in several type 2 
(i.e. absorbed) AGN (e.g. Risaliti \et 2002), but to the best of our 
knowledge not in a NLS1.
\iz\ is a merging and starburst system so there are several sources
of absorption that exist in the galaxy and could feasibly change (e.g. due to 
bulk motion or dissipation of an absorbing component) over the span
of a few years, which separates the two observations.

\subsection{Origin of the Fe~\ka\ variability}
\subsubsection{Flux variations in the 2005 observation}
The broad Fe~\ka\ line in \iz\ shows remarkable short-term variability.
Understanding the origin of this variability is challenging.  

Figure~\ref{fig:cline} suggests that the emission line and continuum variations
could be correlated if we consider the iron-line fluctuations to lag the 
continuum by $10-12\ks$.  Such a delay would place the iron-line emitting region
at approximately $180\rg$ in \iz.  The distance is too large to account for the
width of the iron feature which based on various spectral fits (e.g. Gaussian
profile, disc line profile, blurred ionised reflector), originates within about
$60\rg$.  A second problem is that the amplitude of the variations appears to be larger
in the Fe~\ka\ line than in the $6-10\keV$ continuum that would be producing it.
Even if we assume that the amplitudes of the variations are identical, it would be
unlikely that all of the incident continuum will be reprocessed by the iron producing
region and then emitted back into our line-of-sight.

Another possibility is that the variations are a result of some extrinsic effect
rather than intrinsic changes in the line-emitting region.  For example,  
motion within the accretion disc could account for some of the flux changes
(e.g. Iwasawa \et 2004). 
In this case, one would expect energy modulation in the excess
residual map, which are not present in Figure~\ref{fig:6kevex}.  However,
the strength of the modulations is dependent on the distance 
of the line-emitting region from the black hole and the disc inclination.  It could be that
in \iz\ this combination of parameters does not generate observable modulations. 
 
In a speculative approach, we fit the Fe~\ka\ line light curve 
(Figure~\ref{fig:6kevlc})
with a sine curve.  The fit was statistically good ($\chi^2_{\nu}=1.02$) and
the implied period was $49.1\ks$.  
If the $49.1\ks$ corresponds to the orbital period of the emitting region
around the black hole, then we
can estimate the distance (e.g. following Bardeen \et 1972) to be $\sim25\rg$.
This distance is consistent with that estimated from the width of the broad
Fe~\ka\ line in the spectral models, but it is not clear why only the 
line-emitting region is affected in this manner and not the continuum.

Finally, the Fe~\ka\ variations could be due to variations in some optically-thick
absorber in the line-emitting region.  Such obscuring material could
be created from instabilities in the accretion disc of highly inhomogeneous
flows (e.g. Merloni \et 2006).

\subsubsection{Energy and flux variations in the 2002 observation}
\label{sect:dis02}

The most impressive feature in the 2002 continuum light curve of \iz\ was
a modest X-ray flare concentrated in the $3-12\keV$ band.
The $6.0-6.5\keV$ line-emission is not particularly 
variable and 
demonstrates no connection to the continuum variations.  This behaviour
is typical of other results (e.g. Iwasawa \et 2004).
However, the $5.5-6.0\keV$ line-emission exhibits more intense
variations which mimic the continuum light curve.  In particular, the
peak intensity in the $5.5-6.0\keV$-line light curve 
corresponds well with the peak of the
hard X-ray flare.

Due to limited data, examination of time-resolved spectra during 2002 do 
not yield 
significant
results.  The likely correlation between the continuum and $5.5-6.0\keV$-line 
light
curve suggests that this is emission due to Fe~\ka, which is produced in a
spot on the accretion disc that is illuminated by the X-ray flare (i.e. perhaps
co-rotating with the disc).  The $6.0-6.5\keV$-line emission is representative of the
average broad-line emission, which is produced at distances of $\sim35\rg$ as
suggested by its width.
The $5.5-6.0\keV$-line emission and hard X-ray flare are generated closer to the black hole 
where Doppler and gravitational effects are more intense than the region where 
the $6.0-6.5\keV$ emission line is produced (i.e. $<35\rg$).

It is worth noting that the low-energy emission could be related to the narrow $6.4\keV$
($6\keV$ observed-frame) emission feature that was identified in the 2002 spectrum
by G04.  Since the feature was not detected in the 2005 spectrum
its origin could be connected to the transient $5.5-6.0\keV$-line 
(perhaps the blue-wing of
this line), which is produced during the X-ray flare. 

The combination of a hard X-ray flare, and redshifted and broadened Fe~\ka\ emission
line has been seen
in a very similar object, Nab~0205+024 (Gallo \et 2004c).  However, due to 
limited data, time-resolved spectral modelling did not yield information about
possible line variability.
Re-examination of the Nab~0205+024 data making use of excess residual imaging
may be fruitful.

\subsection{Physical models for \iz}
The level of absorption (neutral and ionised) appears to have changed between
the two epochs.  However, even after considering this there does seem to be
some subtle change in the intrinsic continuum shape.

It is of interest to compare the changes in the context of a physical model
such as reflection and light bending.
In terms of the blurred reflection model, most
of the intrinsic changes can be understood as arising from different relative
contributions of the two continuum components.  At both epochs (and historically), 
\iz\ appears to be in a power law dominated state -- the soft excess is notably 
weak in this object.
 
If the power law component is a compact isotropic emitting source
located above the black hole, as depicted in the light-bending scenario
(e.g. Miniutti \& Fabian 2004), then the changes in the continuum shape 
are largely motivated by the proximity of this source to the black hole.   
In the higher flux 2002 observation, the contribution from the reflection 
component was less, and hence the soft excess was weaker.  The AGN was
brighter and in a power law dominated state.  In 2005, \iz\ was dimmer and
the reflection component (i.e. the soft excess) was stronger.

This is almost certainly a simplified description of the true situation.  One would 
expect other spectral changes to occur as the power law component varies its distance
from the black hole.  For example, with increasing distance the photon index should 
harden (i.e. if due to Comptonisation), 
ionisation parameter may decrease as the accretion disc is less illuminated,
and $R_{in}$ will probably increase.  

As the $0.3-10\keV$ spectrum of \iz\ is dominated by one component and subject
to ionised absorption, it would not be surprising if other physical models
provided adequate fits.
Broad-band absorption models, like the
blurred absorption models (e.g. Gierlinski \& Done 2004), have not been examined
here, but could likely duplicate the spectrum of \iz.  It may be possible
to break this degeneracy by examining the spectral variability, which is complex
in \iz\ and is somewhat unlike what is seen in other NLS1 (Gallo \et 2007). 

\section{Conclusions}

In this work, we presented an $85\ks$ \xmm\
observation and described the mean spectral and timing characteristics of the AGN,
as well as compared the data to a previous, shorter \xmm\ observation from 2002.
The complex ionised absorber is discussed in detail in the work of Costantini \et 
(2007).  The remarkable spectral variability is presented in Gallo \et (2007).
The main findings from this work are the following:

\begin{itemize}
\item[(1)]
\iz\ is $\sim35$ per cent dimmer in 2005 than it was in 2002.
Between the two epochs the level of neutral column density intrinsic 
to \iz\ changed, as well as the ionised absorber. 
In addition, subtle changes in the continuum shape are apparent.  If modelled
with the blurred reflection model, the long-term continuum changes can be
understood as arising from different relative contributions of the two continuum 
components (i.e. direct power law and ionised reflection).

\item[(2)]
Variability is prevalent in all X-ray bands.  The most distinct structure in 
the mean light curve is a flux dip $\sim25\ks$ into the observation.
Average CCFs show a tendency for 
positive lags meaning soft band changes lead harder bands.  
However, examination of 
normalised light curves indicate that the high-energy variations are quite
structured and lags are only apparent in some parts of the light curve.  
Interestingly, the hard X-ray lag first appears at the onset of the flux dip in
the light curve.  This possible alternating lag has been proposed for another
NLS1 (IRAS~13224--3809, Gallo \et 2004b) and it is not consistent with
Comptonisation from a single corona with a uniform temperature and optical depth,
and variable seed photons. 

\item[(3)]
The broad ionised Fe~\ka\ line shows significant variations over the course
of the 2005 observation, which are not clearly due to intrinsic changes in the
Fe~\ka-producing region.  Other, viewer-perspective effects, such as orbital
motion or absorption in the broad iron line-emitting region, 
could be invoked to describe the variability.

\item[(4)]
The broad-line variability during 2002 is re-examined here.
There is evidence suggesting that a relativistically broadened Fe~\ka\
line is produced as a result of a hard X-ray flare illuminating the
inner regions (i.e. $<35\rg$) of the accretion disc.

\end{itemize}


\section*{Acknowledgements}

The \xmm\ project is an ESA Science Mission with instruments
and contributions directly funded by ESA Member States and the
USA (NASA). The \xmm\ project is supported by the
Bundesministerium f\"ur Wirtschaft und Technologie/Deutsches Zentrum
f\"ur Luft- und Raumfahrt (BMWI/DLR, FKZ 50 OX 0001), the Max-Planck
Society and the Heidenhain-Stiftung.
WNB acknowledges support from NASA LTSA grant NAG5-13035 and NASA grant NNG05GR05G.
We thank the anonymous referee for carefully reading the manuscript and 
suggesting improvements.



\bsp
\label{lastpage}
\end{document}